\documentclass[preprint2]{aastex}

\makeatletter
\def\pmb#1{\setbox0=\hbox{#1}%
  \kern-.025em\copy0\kern-\wd0
  \kern.05em\copy0\kern-\wd0
  \kern-.025em\raise.0433em\box0 }

\def\btheta{\pmb{$\theta$}}

\slugcomment{Submitted to ApJ}
\shorttitle{Radio wave scattering in interstellar medium}
\shortauthors{Boldyrev \& K\"onigl}
\begin{document}
\input psfig.sty
\title{Non-Gaussian Radio-Wave Scattering in the Interstellar Medium}
\author{Stanislav Boldyrev and Arieh K\"onigl}
\affil{Department of Astronomy and Astrophysics and Enrico Fermi
Institute, University of Chicago, 5640 S. Ellis Ave., Chicago, IL 60637}
\email{boldyrev@uchicago.edu; arieh@jets.uchicago.edu}

\begin{abstract}
It was recently suggested by Boldyrev \& Gwinn that the
characteristics of radio scintillations from
distant pulsars are best understood if the interstellar
electron-density fluctuations that cause the time
broadening of the radio pulses obey non-Gaussian statistics. In
this picture the density fluctuations are inferred
to be strong on very small scales ($\sim
10^8-10^{10}\,\mbox{cm}$). We argue that such density
structures could correspond to the ionized boundaries
of molecular regions (clouds) and demonstrate that the power-law
distribution of scattering angles that is required to match the
observations arises naturally from the expected intersections of
our line of sight with randomly distributed, thin, approximately
spherical ionized shells of this type. We show that the observed 
change in the time-broadening behavior for 
pulsar dispersion measures $\lesssim 30\, {\rm pc}\, {\rm cm}^{-3}$
is consistent with the expected effect of the general ISM turbulence, which
should dominate the scattering for nearby pulsars. 
We also point out that if the clouds are
ionized by nearby stars, then their boundaries may become
turbulent on account of an ionization front instability. This
turbulence could be an alternative cause of the inferred density
structures. An additional effect that might contribute to the
strength of the small-scale fluctuations in this case is the
expected flattening of the turbulent density
spectrum when the eddy sizes approach the proton gyroscale.
\end{abstract}

\section{Introduction}
\label{introduction}
Radio signals received from distant pulsars fluctuate in 
time and space due to inhomogeneities in the ionized component of
the interstellar medium. Observations of pulsar 
scintillations have thus served as a valuable tool for reconstructing 
the statistics  of interstellar electron-density fluctuations
\citep[e.g.,][]{rickett,armstrong,scalo}. Among a variety of
observational quantities, special attention has been given to
the time shape of the pulsar signal $I(\tau)$ (the signal
intensity as a function of time). This function arises from the
interference of waves propagating along different paths (similar
to the Feynman interpretation of the propagation of a quantum
particle). The deviation of the wave paths from a straight line
causes a time-broadening of the arriving signal. The intensity
function $I(\tau)$ can be interpreted as the  
probability density function (PDF) of the time delays caused by
different paths of propagation. Since the statistics of the path
deviations are directly related to the statistics of
electron density fluctuations, the function $I(\tau)$ provides
information on the latter.

\citet{boldyrev1,boldyrev2,boldyrev3} recently used the observed
time shapes of the measured pulses to probe the shape of 
the distribution function of the underlying density fluctuations.
Based on a comparison of analytical results and observational data, they
proposed that the density fluctuations responsible for the pulse broadening 
are non-Gaussian, with the distribution function of the
resulting angular path deviations
having a slowly declining, power-law asymptotic. They further
deduced that the density fluctuations should be strong on very small 
scales (estimated in \S~\ref{difference} to be $\sim 10^8-10^{10}\,\mbox{cm}$).

After being suitably rescaled by the pulsar distance 
and the observation wavelength, the observed pulse time shapes 
exhibit a somewhat ``universal'' behavior \citep[e.g.,][]{bhat,lohmer}, which
indicates that similar processes govern the density fluctuations
along different lines of sight. Two intriguing questions then
arise regarding the physics of interstellar plasma
fluctuations. First, how are strong small-scale density
fluctuations generated? And second, how can density fluctuations
produce a non-Gaussian, power-law distribution of scattering angles?

In addressing these questions one can consider two 
distinct (although not mutually exclusive) possibilities. In one
interpretation the observed universality is attributed to a
general {\em statistical} property of density fluctuations. 
For example, in Kolmogorov-type turbulence, small-scale fluctuations arise 
as a result of an energy cascade from large to small scales. The
properties of the observed signals might then be linked to the
universal properties of such a cascade. The second possibility
is that density fluctuations are strongly nonuniform and
spatially intermittent (e.g., clumps, filaments, and shocks).
In this case, the universality could reflect a
certain inherent property of the 
density structures. This property should be fairly robust in
view of the fact that
different mechanisms (not necessarily related to turbulence) 
could in principle give rise to a particular structure.
For example, scattering by shocks may be universal if the only property that 
matters is a sharp density 
discontinuity~\cite[see, e.g.,][]{boldyrev3} .

Let us briefly discuss the first possibility. We assume that
interstellar scattering occurs in turbulent ionized regions (HII
regions) of mean density $n_2\sim 10^2\,\mbox{cm}^{-3}$.  Such
regions appear when bright stars turn on within or in the
vicinity of cold molecular clouds
\cite[e.g.,][]{dyson}. Large-scale density fluctuations in a
cold molecular cloud of mean density $n_0\sim
10^2\,\mbox{cm}^{-3}$ are of the order $\Delta n_0\sim
10^2\,\mbox{cm}^{-3}$, since such clouds are usually turbulent
with Mach numbers greater than 1~\citep{larson}.  When cold
neutral gas in a molecular cloud is ionized by a newborn massive
star emitting $\sim 10^{49}$ ionizing photons~$s^{-1}$, the
radius of the ionized sphere (the Str\"omgren radius) is
$R_s\sim 10^{19}\,\mbox{cm}$. This ionization happens very fast,
during the first $10^4$ years after the radiation turns on. This
time interval is much shorter than the sound crossing time in
the ionized gas within the Str\"omgren sphere, where the sound
speed is $C_2\sim 10^6\,\mbox{cm\, s}^{-1}$. Subsequently, the
initial density inhomogeneities relax due to pressure gradients,
which drive turbulent internal motions in the HII region. The
longest relaxation time corresponds to the largest spatial
scale, and therefore the turbulence that survives over the
lifetime ($\sim 10^6\, {\rm yr}$) of the star is effectively
stirred on the scale $l_{\rm out}\sim R_s$, where the velocity
and density fluctuations are of the order of $C_2$ and $\Delta
n_0$, respectively.

The interstellar medium is magnetized and the magnetic field is
likely to play a role in any turbulent energy cascade. 
As discussed by several authors
\citep[e.g.,][]{higdon,goldreich,lithwick}, density
perturbations in an MHD turbulence are associated with
compressible and entropy
modes and are {\em passively} advected by the Alfv\'enic cascade
toward the smallest (dissipative) spatial scales. The cascade proceeds
predominantly in the direction perpendicular to the magnetic field
and continues until the fluctuations reach the 
scale of the ion gyroradius~$\rho_i$. 
A typical strength of the
magnetic field in the Galaxy is a few $\mu{\rm G}$ 
\citep{zweibel}. For an ion temperature
$T_i\sim 8000\, \mbox{K}$ and magnetic field amplitude 
$B\sim 3\, {\mu}\mbox{G}$,  
the gyroradius is $\rho_i\sim 3\times
10^7\,\mbox{cm}$, so the density fluctuations can reach very small scales.

However, a problem arises when we address the statistics of
density fluctuations that are advected as a passive scalar in
MHD turbulence. Numerical results show that passive scalar
fluctuations in an incompressible turbulence have an exponential
(or stretched exponential) distribution
\citep[e.g.,][]{warhaft}, not the power-law distribution that we
are trying to explain.  Another problem arises when we recall
that the fluctuations were inferred to be strong on scales as
small as $\sim 10^8-10^{10}\,\mbox{cm}$. To reach these scales,
the envisioned cascade would have to pass through a scale where
the eddy turnover time is comparable to the radiative cooling
time of the plasma, at which point the density fluctuation may
be strongly damped \citep{lithwick}. The radiative cooling time
for ionized gas of temperature $\sim 8000\,\mbox{K}$ is given by
$t_{\rm cool}\sim 6\times 10^{11}n_{2}^{-1}\,\mbox{s}$
\citep{spitzer}, so for $n_2\sim 10^2\,\mbox{cm}^{-3}$ and a
Kolmogorov-like turbulence, the cooling scale is $l_{\rm
cool}\sim 10^{14}\,\mbox{cm}$, significantly larger than the
scales we are trying to account for.

In view of the foregoing arguments it appears difficult to attribute
the inferred non-Gaussian scattering by small-scale
density fluctuations to a nearly incompressible MHD turbulence
excited on scales $l_{\rm out} \sim 10\, {\rm pc}$. In
\S~\ref{origin} we discuss a possible means of exciting
the turbulence on scales smaller than $l_{\rm cool}$ (to account
for the inferred strength of the small-scale fluctuations) and
appeal to kinetic effects (which could make the turbulence
compressible) for further enhancement of the fluctuations and
the possible establishment of a power-law spectrum on scales
close to $\rho_i$. However, our main focus in this paper is on
exploring the second of the above two possibilities for the
origin of the inferred density fluctuations, namely, that the
spatial distribution of the scattering electrons is highly
intermittent. Specifically, we propose that these electrons
are contained in thin, shell-like structures. Although we show
that the statistical properties of the scattered radio pulses are
independent of the physical origin of the shells, we argue that
the latter can be most plausibly identified with the thin
surface layers of molecular regions (clouds) 
that become ionized by the
interstellar radiation field or by ionizing photons from nearby
stars.

The ionized surfaces of dense molecular clouds were previously
invoked by \citet{yusef-zadeh} to explain (on the assumption
that these layers are magnetized and turbulent) the observed
anisotropy in the angular broadening of the radio source Sgr A*
in the Galactic center. Furthermore, localized scattering
structures (such as shocks or filaments) have been invoked by
\citet{lambert} and by \citet{cordes} to explain the observed
scaling of pulsar signals with frequency and the results of
certain diffractive measurements. Small-scale localized electron density 
structures have also been inferred from observations of 
extreme scattering events \citep[e.g.,][]{lazio}, pulsar arclet 
structures \citep[e.g.,][]{hill}, and pulsar 
scintillations \citep[e.g.,][]{cordes2}. Although these investigations
were carried out in the context of the Gaussian model and thus do
not bear directly on the results presented here (which aim to
justify a non-Gaussian interpretation), their appeal to sharply
delineated scattering regions  
has inspired our work.

In this paper we demonstrate that the non-Gaussian, power-law
statistics inferred for the angular scattering of pulsar radio
pulses need not be related to the character of density
fluctuations inside turbulent scattering regions but can be
fully explained in terms of the spatially intermittent nature of thin
ionized shells in the interstellar medium (which in principle
need not even be turbulent). Using purely geometric-optics considerations, 
we show that scattering by thin spherical shells produces a
distribution of angular deviations $\Delta \theta$
with an asymptotic power-law probability density $P(\Delta
\theta)\propto \Delta \theta^{-5/3}$. Interestingly,
this is precisely the form deduced in \cite{boldyrev1} by
applying the non-Gaussian (L\'evy) scattering theory to the
pulsar scintillation data. This correspondence strongly suggests
that the scattering regions that produce refractive radio
scintillations have a morphology of curved shells.

After briefly outlining the theory of pulsar scintillations
(\S~\ref{motivation}), we discuss the key differences between
Gaussian and non-Gaussian models of density fluctuations
(\S~\ref{difference}). In \S~\ref{shells} we flesh out our proposal
that scattering by thin ionized boundaries of molecular clouds  
can naturally explain the inferred non-Gaussian statistics of
the scintillations. In \S~\ref{origin} we consider the
case in which  the ionizing radiation is strong enough to trigger an
ionization front instability and possibly the development of
turbulence within the ionized cloud boundaries. Our results are
summarized in \S~\ref{conclusion}.

\section{Pulsar Scintillations and Interstellar Density Fluctuations}
\label{motivation}

In this section we explain how the statistics of electron-density 
fluctuations can be diagnosed by observations of the time shapes of
pulsar signals. To this end, we assume that the scattering (refraction) 
of a radio wave coming from a distant pulsar 
occurs in separate planar phase screens, uniformly 
placed along the line of sight to the pulsar. We denote the
line-of-sight axis by~$z$ and the coordinates in the
perpendicular plane by a two-dimensional vector~${\bf y}$. We
also assume that the scattered wave is planar and propagates close to 
the line of sight. 
To analyze this model one can make use of the so-called parabolic 
approximation, which assumes that the wave amplitude changes
slowly compared to its phase. The details of this analysis can
be found in a number of references
\citep[e.g.,][]{tatarskii,uscinski,williamson72,lee,rickett,boldyrev3};
here we only give a qualitative discussion of the results. 

Each scattering screen generates phase fluctuations that cause the 
transmitted wave to contain many different angular components in its
spectrum rather than follow a single direction. The spatial scale of
the fluctuations that contribute to an angular deviation $\Delta \theta$ 
(expressed in radians) is $y\sim \lambda/\Delta \theta$, where $\lambda$ 
is the wavelength. The difference of phases accumulated in a
single screen along two lines of sight separated by the distance
${\bf y}$ is given by $\Delta \Phi=\lambda r_0{\Delta {\cal{N}}}({\bf
y}) $. It is proportional to the
line-of-sight integrated density difference
\begin{eqnarray}
{\Delta {\cal{N}}}({\bf y})=
\int\limits^{l_0}_0\left[n({\bf y}_1, z)-n({\bf y}_2, z)\right]
d z\ ,
\label{n}
\end{eqnarray}
where ${\bf y} \equiv {\bf y}_1-{\bf y}_2$ and $r_0\equiv e^2/(m_e
c^2)=2.82\times 10^{-13}\, {\rm cm}$ is the classical 
electron radius. The integration 
length~$l_0$ should exceed the characteristic size of the 
scattering region. To produce an angular
deviation $\Delta \theta$, the phase difference should be 
$\Delta \Phi \sim 2\pi$ on the scale $y\sim \lambda/\Delta \theta$.
(where here, as in the rest of this paper, angles are measured in radians).

The signal emitted by a pulsar has a rather 
short duration (several percent of the pulsar period). The
received signal is broader, with its intensity~$I(\tau)$
exhibiting a sharp rise and a slow decline. This happens because
waves propagating along different paths from the pulsar to Earth experience
different angular deviations by the electron density prisms
that intercept them. This, in turn, leads to different time delays
relative to straight (undeflected) propagation. The signal 
intensity corresponding to a time delay~$\tau$ is proportional 
to the probability for a path to be delayed by~$\tau$
\citep[e.g.,][]{boldyrev3}.

Assume that the number of scattering screens is~$N$ and the distance 
between two neighboring screens along the line of sight is~$z_0$, so 
that $d=Nz_0$ is the 
distance to the pulsar. The angle of the path
in the segment between screens $m-1$ and $m$ is 
a sum of the angular increments accumulated at the preceding segments, 
$\btheta_m=\sum_{s=1}^{m}\Delta \btheta_s$. 
In general, $\Delta \btheta$ is 
a two-dimensional vector, pointing in the direction of the path 
deviation.  The time delay introduced by 
the $m$th scattering segment is therefore 
$\Delta \tau_m=(z_0/c)(1-\cos\theta_m)\sim (z_0/2c)\btheta^2_m$, 
and the total time delay is the sum of the individual delays,
\begin{eqnarray}
\tau=\sum\limits_{m=1}^N\Delta \tau_m=\frac{z_0}{2c}
\sum\limits^N_{m=1}\left[\sum\limits_{s=1}^m \Delta \btheta_s \right]^2\ .
\label{tau}
\end{eqnarray}
In this equation the individual angular increments $\Delta \btheta$ are 
independently and identically distributed two-dimensional vectors. The 
intensity $I(\tau)$ of the received signal, which is effectively 
the PDF of $\tau$, can be found numerically 
using equation~(\ref{tau}) once the distribution function 
for the vectors $\Delta \btheta_s$ is known.

In the classical 
theory of scintillations the vectors $\Delta \btheta_s$ are
assumed to have an independent and identical 
Gaussian distribution, in which case an exact
analytic solution for the $\tau$-distribution function $I(\tau)$ can be derived 
\citep[e.g.,][]{uscinski,williamson72}. 
However, the scaling and shapes of pulse 
profiles predicted by this theory seem to disagree with 
the observational results for distant pulsars, as first noted by
\citet{sutton} and \citet{williamson}. In a recently proposed
alternative model, \citet{boldyrev1} 
demonstrated that the observed profiles can be matched if the individual 
fluctuations $\Delta \btheta_s$ have a non-Gaussian, power-law--like
declining distribution. In this case the wave path angle~$\theta_m$ exhibits a
so-called L\'evy flight rather than the standard random walk.

The L\'evy distribution function  
is a general distribution function whose
convolution with itself produces the same function again
(appropriately rescaled). 
The Gaussian distribution is a
particular case of a L\'evy distribution.  
A detailed discussion of L\'evy distributions can be found 
in \citet{klafter}; the application to  wave propagation in the
interstellar medium is considered in \citet{boldyrev1,boldyrev2,boldyrev3}. For our
present purposes, the main relevant property of the L\'evy distribution 
is its slow (power-law) asymptotic decline. Whereas
$\theta_m^2\propto m$ for a Gaussian random walk, the scaling
produced by a L\'evy flight 
is $\theta_m^2\propto m^{2/\beta}$, where $\beta$ (which
satisfies $0<\beta \leq 2$) is the parameter of the 
L\'evy distribution. For $\beta < 2$, 
the L\'evy PDF exhibits a power-law decline, 
$P_{\beta}(\Delta\theta)\propto |\Delta\theta|^{-1-\beta}$ for 
$|\Delta\theta| \gg \Delta\theta_0$, 
where $\Delta\theta_0$ is a typical angular fluctuation value. 

Pulsar signals are observed to satisfy the scaling $\tau \propto
N^4$ (more precisely, the observationally inferred scaling is $\tau\propto DM^4$, 
where the pulsar dispersion measure, $DM$, is proportional to 
the distance to the pulsar and, therefore, to the number of 
scattering events,~$N$), which motivated \citet{boldyrev1,boldyrev2,boldyrev3} to
propose that radio scintillations are described by a L\'evy
model with $\beta\sim 2/3$. A Gaussian distribution 
corresponds to $\beta=2$ and implies a linear scaling of $\theta_m^2$ with $m$,
which is not consistent with the observational data. In \S ~3 we demonstrate that the difference in the implied scaling
between these two models corresponds to a profound difference in the physics 
of the respective scattering mechanisms.

\section{L\'evy Statistics of Scintillations}
\label{difference}
We assume that the PDF of $\Delta\theta$ has a 
characteristic width $\Delta\theta_0$ (corresponding, say,
to the $1/e$ amplitude level), and we denote the width of the
resulting PDF of $\tau$ by $\tau_0$. Equation~(\ref{tau}) and
the scaling $\theta_m^2 \propto m$ imply that $\tau_0$ scales as
$(\Delta \theta_0)^2 N^2$ in a Gaussian model. On the other hand, if we 
consider $\tau_0$ to arise from the action of individual
fluctuations, each of magnitude $\Delta\theta$, we find (using 
eq.~[\ref{tau}] again) that $\tau_0\propto (\Delta\theta)^2N^3$. By
comparing these two expressions we obtain, as expected, $\Delta\theta\sim
\Delta\theta_0/N^{1/2}$, which is consistent with the notion
that the typical time delay may be produced by many small
deflections.

The situation is quite different in the L\'evy case with $\beta=2/3<1$.  
Equation~(\ref{tau}) implies $\tau_0 \propto (\Delta \theta_0)^2 N^4$, 
and if we compare this expression with our estimate in terms of
individual fluctuations, $\tau_0\propto (\Delta\theta)^2 N^3$,
we infer that such fluctuations must be {\em
large},~$\Delta\theta \sim N^{1/2} \Delta \theta_0$. 
But such large individual 
fluctuations have a small probability, 
$P\propto 1/(\Delta \theta)^{1+\beta}\propto 1/N^{5/6}$,
and it is therefore highly unlikely that all $N$ individual
fluctuations are of that order. Instead, the most probable situation
is that {\em one} of the fluctuations is extremely large, 
$\Delta\theta =\Delta\theta_{max} \sim N^{3/2}\Delta\theta_0$, 
whereas all the others are 
small ($\lesssim \Delta\theta_0$). Thus, in each particular 
realization, the total scattering is most probably produced
by only one, randomly chosen, screen. This behavior 
is a general property of sums of L\'evy-distributed variables 
with $\beta <1$ \citep{feller}. When the pulse shape is
averaged over a certain time interval (or over a statistical
ensemble), these screens can 
be different in different realizations, since the interstellar medium 
fluctuates and both 
the pulsar and the Earth move through it. 

We can now estimate the width of the pulse as $\tau_0\sim z_0
N(\Delta\theta_{\rm max})^2/c \sim d(\Delta\theta_{\rm max})^2/c$, where
$d=Nz_0$ is again the pulsar distance. This expression has a geometric
meaning, namely, that only one refractive event dominates the path
deviation from a straight line. For the typical parameters of distant
pulsars  \citep[e.g.,][]{bhat}, $d\sim 3 \times 10^{22}\,\mbox{cm}$, 
$\tau_0\sim 0.01\,\mbox{s}$,  and $\lambda \sim
30\, \mbox{cm}$ (corresponding to a frequency of $1\, {\rm GHz}$), we
obtain $\Delta\theta_{max}\sim 10^{-7}\,{\rm rad}$ and a 
density fluctuation scale
$y\sim \lambda/\Delta\theta_{max}\sim 3\times  10^8\, \mbox{cm}$.  In
view of the scaling $\tau\propto d^{4}$, closer pulsars (characterized
by $d\sim 3\times 10^{21}\,\mbox{cm}$ and $\tau_0\sim  10^{-6}\,\mbox{s}$) 
correspond to $\Delta \theta_{max} \sim 3\times 10^{-9}\,{\rm rad}$ and 
$y\sim 10^{10}\,\mbox{cm}$.

What regions in the interstellar medium might be responsible 
for the indicated strong scattering events? To address this question,
we first consider whether the maximal angular deviation estimated 
above could be produced by regions of uniform and homogeneous 
turbulence. In the following discussion we therefore neglect the contribution of
the boundaries of the turbulent regions or of any other structures.

The amplitude of the
density fluctuations that give rise to the large inferred
angular deviations can be estimated from the results presented
after equation~(\ref{n}), which imply
\begin{eqnarray}
\Delta {\cal{N}}(y)/y\sim \Delta \theta_{max}/(\lambda^2 r_0)\sim 10^3\,
\mbox{cm}^{-3}\ . 
\label{nperp}
\end{eqnarray}
This defines the value that the density fluctuations must be able to reach on 
the scale~$y$. Suppose that the density fluctuations are
associated with a turbulent cascade. If the density is passively
advected by the fluid turbulence \citep[e.g.,][]{lithwick} then
the density scaling follows that of the velocity field, 
and for a spatially homogeneous turbulence with an outer
scale~$l_{\rm out}$ we have $\delta n(l)/\Delta n_0\sim \delta
v(l)/\Delta v_0\sim (l/l_{\rm out})^{\alpha/2}$. 
Here $\Delta v_0$ and $\Delta n_0$ are, respectively, the rms values of
the velocity and density fluctuations on the scale $l_{\rm out}$  and $\alpha$ is
the turbulence scaling exponent, which ranges between 2/3 (the Kolmogorov case)
and~2.  
This estimate neglects damping effects, which can be different
for $\delta n$ and $\delta v$; we discuss damping further on in
this section.

As we pointed out in \S~\ref{introduction}, the 
distribution function of density fluctuation amplitudes in
Kolmogorov turbulence declines  faster than a power law. The central limit 
theorem then implies that the distribution of line-of-sight integrated density 
fluctuations is Gaussian. Under these conditions, 
the $y$-dependence of the integrated density difference is
\begin{eqnarray}
\Delta {\cal{N}}(y) \sim l_0^{1/2}l_{\rm out}^{1/2} \Delta n_0 (y/l_{\rm 
out})^{\delta }\ ,
\label{calN}
\end{eqnarray} 
where $\delta=(1+\alpha)/2$ for $\alpha<1$  and $\delta =1$ 
for $\alpha\geq 1$. 
Equation (\ref{calN}) follows directly from the expression for the 
correlator
$\langle[\Delta {\cal{N}}(y)]^2 \rangle$ in a homogeneous and
isotropic turbulence. 
Note that this correlator is proportional to the integration
distance $l_0$, as expected for Gaussian fluctuations.

In the standard picture of scintillations (the first scenario discussed
in \S~\ref{introduction}), turbulent HII regions around 
bright stars are considered to be the likely
sites of radio-wave scattering. In this case the outer scale of the
turbulence, $l_{\rm out}\sim 10^{19}\,\mbox{cm}$, is of the
order of the size $l_0$ of the Str\"omgren sphere. Setting 
$l_0\sim l_{out}$ in equation~(\ref{calN}) and comparing with
equation~(\ref{nperp}), we obtain the following condition on the
large-scale density fluctuations:
\begin{eqnarray} 
\Delta n_0 \sim 10^3 \,(y/l_{\rm
out})^{(1-\alpha)/2}\,\mbox{cm}^{-3}\ . 
\label{dn} 
\end{eqnarray}
Taking the velocity difference at $l_{\rm out}$ to be of the
order of the sound speed in the ionized region ($C_2\sim
10^6\,\mbox{cm/s}$) and assuming (conservatively) that the
turbulence in the H II region indeed obeys a Kolmogorov scaling
($\alpha=2/3$) and that $y \sim 10^8\,\mbox{cm}$, we find that
equation~(\ref{dn}) is satisfied on this scale for electron density
fluctuations $\Delta n_0 \gtrsim 30\, \mbox{cm}^{-3}$.

Although this value is consistent with the typical densities of
molecular clouds, in reality the required value of $\Delta n_0$
should be  much larger (and therefore implausible) in view of the
fact that the compressible and entropy MHD modes associated with the density
fluctuations are damped when the wave cascade 
passes through the cooling scale \citep[see][]{lithwick}, which
is about $10^{14}\,\mbox{cm}$ in this case (see \S~\ref{introduction}).

The cooling length constraint limits the ability of the standard
picture to explain
the strong, small-scale density fluctuations that are required in the
L\'evy-flight scenario.  Furthermore, as discussed in
\S~\ref{introduction}, the distribution of passive scalar fluctuations
in the incompressible MHD model is not power-law--like. These two
difficulties suggest that non-Gaussian scattering cannot be 
explained within the standard framework.  In \S~\ref{shells} we
demonstrate that such scattering may be explained if it is produced in
thin shells --- the ionized boundaries of molecular regions (clouds).

\section{Non-Gaussian Scattering by Ionized Boundaries of Molecular
Clouds} 
\label{shells} 
A typical value for the photon flux of ionizing
interstellar radiation ($h\nu > 13.6\,\mbox{eV}$) through a unit surface
can be estimated by converting the standard Habing flux \citep{habing}
of dissociating photons ($6<h\nu < 13.6\,\mbox{eV} $) using a blackbody
spectrum of temperature $T\sim 3\times 10^4\,\mbox{K}$
\citep[see][]{yusef-zadeh}.  This gives $J_0\sim 2\times
10^7\,\mbox{cm}^{-2}\,{\rm s}^{-1}$. In the vicinity of a bright star the
ionizing flux can be several orders of magnitude higher. For example, at
a distance of $1\,\mbox{pc}$ from a star emitting $S=10^{49}$ ionizing
photons $s^{-1}$, this flux is $J\sim 10^{11}\,\mbox{cm}^{-2}\,{\rm s}^{-1}$.
The mean free path of an ionizing photon that penetrates a neutral
medium of density $n_0$ is $\lambda_{\rm ph}\sim 1/(\alpha_0 n_0)$, where
$\alpha_0=6.8\times 10^{-18}\,\mbox{cm}^2$ is the hydrogen ionization
cross section at the threshold energy $h\nu=13.6\,\mbox{eV}$
\citep[e.g.,][]{spitzer}.

When a homogeneous molecular cloud is irradiated by an ionizing photon
flux~$J$ it develops an ionized ``skin'' whose width $\Delta r_i$ can be
found from the ionization balance condition $J=\beta_2 n_0^2\Delta r_i$,
where $\beta_2$ is the recombination coefficient ($\beta_2\sim
2\times 10^{-13}\,\mbox{cm}^3\mbox{s}^{-1}$ for $T\sim 10^4\,\mbox{K}$;
e.g., \citealt{dyson}). For $n_0\sim 10^2\,\mbox{cm}^{-3}$ and $J=J_0$ one
finds $\Delta r_i\sim 10^{16}\,\mbox{cm}$, which can be taken as the
thickness of the ionized boundary. The width of the transition layer
between the ionized skin and the neutral interior of the cloud is of the
order of $\lambda_{\rm ph}$ ($\approx 10^{15}\,\mbox{cm}$ for the
adopted value of $n_0$).
{
\begin{figure} [tbp]
\centerline{\psfig{file=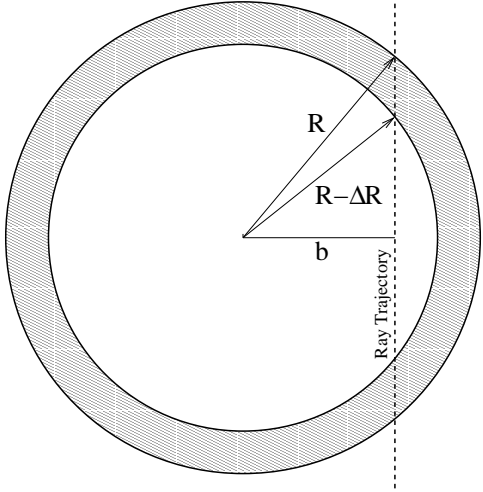,width=2.5in,angle=0}}
\caption{Scattering by a thin spherical shell. The ray's angular deviation 
is very small, so its trajectory is well approximated by a straight line.
}
\label{shell}
\end{figure}
}

We now demonstrate that thin scattering shells (representing the ionized
boundaries of molecular clouds) can produce non-Gaussian scintillations 
even if they
are not turbulent. In particular, we show that scattering by thin curved
layers (which we represent for simplicity as thin spherical shells)
produces a power-law distribution of the scattering angles. Most
remarkably, the derived power-law index coincides with the value
originally inferred by \citet{boldyrev1} from an interpretation of the
scintillation data in terms of the L\'evy theory (see \S~\ref{motivation} and \ref{difference}).
The agreement with the observations suggests that a shell-like structure
is a general feature of the ionized scattering regions in the
interstellar medium.

To derive these results, we assume that each of the electron-scattering
regions has the form of a thin spherical shell of radius~$R$ 
and thickness~$\Delta R\ll R$ (see Fig.~\ref{shell}). 
We denote by $\Delta n$ the electron-density excess 
in the shell relative to its surroundings and employ geometric 
optics to calculate the angular 
deviation $\Delta \theta$ experienced by a ray that intercepts the shell
at a distance~$b$ from the center. 
For a typical density excess $\Delta n\sim 10^2\,\mbox{cm}^{-3}$ this
deviation is very small, $\Delta \theta \ll 1$, so the trajectory of the
ray through the scattering region can be approximated by a straight line. 

Scattering by a shell may be viewed as a superposition of 
scatterings by two spheres of radii~$R$ and~$R-\Delta R$,
respectively. The total angular deviation produced by a sphere of radius $r$
is readily obtained using Snell's law and is given as a function of the impact 
parameter $b$ by
$\theta_r(b)=\Delta \theta_s (b/r)(1-b^2/r^2)^{-1/2}$. In this
expression $\Delta \theta_s\equiv\lambda^2 r_0 \Delta n/\pi$, where $\lambda$ 
and $r_0$ are again the wavelength and the classical electron radius,
respectively.

The angular deviation produced by the shell is $\Delta \theta(b)= 
\theta_R(b)- \theta_{R-\Delta R}(b)$. It is given,
up to first order in the small parameter~$\Delta R/R$,
by $\Delta \theta(b)=\Delta \theta_s (b\Delta R/R^2)(1-b^2/R^2)^{-3/2}$. 
Large angular deviations, $\Delta \theta (b) \gg \Delta \theta_s \Delta R/R$, 
correspond to $b\approx R$, and in this limit we obtain
\begin{eqnarray} 
\Delta \theta(b) \approx \Delta \theta_s (\Delta R/R) (2\Delta
b/R)^{-3/2}\ , 
\label{deltatheta}
\end{eqnarray}
where $\Delta b\equiv R-b$. 
When $b$ is not close to~$R$, the angular deviations are small, 
$\Delta \theta \sim 
\Delta \theta_s (\Delta R/R)$.

The probability density for the ray to pass at a distance~$b$ from 
the center is estimated using $P(b)\mbox{d}b=2\pi b\mbox{d}b/(\pi R^2)
\sim (2/R) \mbox{d}b$, where we again assume~$b\approx R$.  The
corresponding PDF of the deviation
angle~$\Delta \theta$ can be found using equation~(\ref{deltatheta}),
which gives
\begin{equation}
\label{P_Dtheta}
P(\Delta \theta)\sim (2/3)(\Delta \theta_s \Delta R/R)^{2/3}
(\Delta \theta)^{-5/3}\ .
\end{equation}
This expression describes the asymptotic
behavior of the PDF for $\Delta \theta \gg \Delta \theta_s \Delta
R/R$. The derived exponent~($-5/3$) of the power-law decline of
$P(\Delta \theta)$ coincides with the exponent predicted in the
L\'evy model~(see \S~\ref{motivation}).
Overall, the PDF $P(\Delta \theta)$ has a typical width of order
$\Delta \theta_0 \sim \Delta \theta_s (\Delta R/R)$
(corresponding to impact parameters
$b$ not close to $R$). At larger angular deviations the PDF develops the
power-law shape obtained above, which extends up to the cutoff 
value $\Delta \theta_{c} \sim \Delta \theta_s (R/\Delta R)^{1/2}$
(attained when $\Delta b$ decreases below  $\sim \Delta R$). 
For even larger angular deviations (i.e., for $\Delta b \ll \Delta R$) 
the scattering is produced only by the outer sphere in Figure~\ref{shell}
and the PDF exhibits a fast decline, $P(\Delta \theta)\propto (\Delta
\theta)^{-3}$.  This PDF is sketched in Figure~\ref{pdf}.
{
\begin{figure} [tbp]
\centerline{\psfig{file=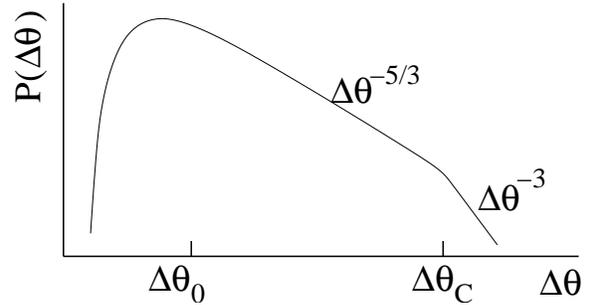,width=3.1in,angle=0}}
\caption{A sketch of the angular-deviation PDF on a log-log scale.  
}
\label{pdf}
\end{figure}
}
  
For our fiducial parameter values $\Delta \theta_s\sim 10^{-8}\,{\rm rad}$
and $\Delta \theta_{max} \sim 10^{-7}\,{\rm rad}$ (see
\S~\ref{difference}). Based on equation (\ref{deltatheta}), the
inferred value of the maximal angular deviation is attained when
$(\Delta R/R)(2\Delta b/R)^{-3/2}\sim \Delta \theta_{max}/\Delta \theta_s 
\sim 10$.  
Given that $\Delta b$ cannot be smaller than the shell thickness (i.e., $\Delta
b\gtrsim \Delta R$), the shell thickness must satisfy~$\Delta
R/R<10^{-2}$. If the shell's radius of curvature is $R\sim
10\,\mbox{pc}$, we have~$\Delta R < 3\times 10^{16}\,\mbox{cm}$, which
is in excellent agreement with our estimate of $\Delta r_i$ in
typical molecular clouds. We note, however, that to calculate
the total probability distribution function of the angular
deviation $\Delta\theta$ one needs to convolve the
distribution $P(\Delta \theta)$ 
(which was obtained for a boundary of given curvature and density) 
with the curvature radii and density distributions of real clouds. This could in principle
change the value of the coefficient $(\Delta \theta_s \Delta
R/R)^{2/3}$ in equation (\ref{P_Dtheta}) as well as the range of
angular deviations over which the derived power-law asymptotic
applies. The curvature radii of a cloud's boundary are typically
smaller than the 
cloud size, so we may use the cloud size as an upper bound on the 
curvature~$R$. Now, $\Delta \theta_s \Delta R/R$
scales approximately as $1/(n_0 R)$. To the extent
that the column density of molecular clouds is approximately
a constant \citep[e.g.,][]{larson,mckee}, the coefficient in
equation (\ref{P_Dtheta}) would change little from cloud to cloud. In
a similar vein, the cutoff angular deviation $\Delta\theta_c$
scales approximately as $R^{-3/2}$. Therefore, our choice of fiducial cloud
radius ($R=10\, {\rm pc}$), which lies near the upper end of the observed
distribution \citep[e.g.,][]{spitzer}, is conservative in that
it leads to a lower bound on $\Delta\theta_c$. All in all,
we expect that integrating the expression (\ref{P_Dtheta}) over
a cloud distribution would have little effect on the derived
asymptotic form of $P(\Delta \theta)$ and, in particular, would
not change the inferred value of its power-law exponent.

It is interesting to consider some observational implications of 
the proposed model.  
As we established in this 
section, $\Delta R/R< (\Delta \theta_s/\Delta \theta_{max})^2$. 
On the other hand, from \S~\ref{difference} we know that 
when the scattering angles have a L\'evy distribution with $\beta=2/3$, 
the number of scattering structures along the line of sight scales 
with the maximal deflection angle as  
$N^{3/2}\sim \Delta \theta_{max}/\Delta \theta_0\sim 
(\Delta \theta_{max}/\Delta \theta_s)(R/\Delta R) $. 
Therefore, our considerations suggest that the number of
scattering structures out to the characteristic distance $d\sim
3\times 10^{22}\,\mbox{cm}$ can be estimated as
$N> (\Delta \theta_{max}/\Delta \theta_s)^2\sim 100$, or about 
one scattering structure per $100\,\mbox{pc}$. 
This estimate is consistent with the overlap radius of radiatively
cooled supernova shells (within which dense clouds are envisioned to
form) in the \citet{ostriker} model of the interstellar medium. The estimated
number of scattering regions could in principle be smaller
(resulting in a weaker constraint on their separation) 
since several independent scattering structures (ionized boundaries) might in
practice form inside a single scattering region. 

The physical basis for the statistical properties of the
thin, curved scattering structures in the proposed interpretation is their
intermittent spatial distribution. Specifically, 
the free electrons responsible for 
the scattering are envisioned to occupy only a
small fraction of the interstellar medium --- namely, the ionized boundaries
of molecular clouds. Furthermore, the molecular clouds 
themselves are sparsely distributed in space, so the effects of two 
different clouds on wave propagation are uncorrelated. 
Thus, a radio signal from a distant pulsar does not 
experience significant scattering over most of its propagation distance
to Earth.  However, when its trajectory intersects a curved, ionized
cloud boundary, the scattering may be strong. As discussed
in~\S~\ref{difference}, this behavior is characteristic of L\'evy-flight
wave paths. The spatial intermittency in this scenario is distinct
from the intermittent nature of turbulence that underlies the
deviation angle statistics in the traditional
interpretation of interstellar scintillations. In contrast with the 
traditional picture, the cumulative effect of many independent 
scatterings by different thin shells is not Gaussian (which can
be understood formally from the fact that the central 
limit theorem does not apply to slowly declining distributions);
it is instead a L\'evy flight with $\beta\sim 2/3$, as implied by the
observations (see \S~\ref{motivation}).
 
The foregoing derivation of the statistical properties of
scattering by thin shells is based exclusively on geometric 
considerations. If, however, the shells are turbulent, then the
effects of turbulence need to be considered in conjunction with
those of geometry. Turbulence in the ionized boundary of a molecular cloud 
can be a natural consequence of the turbulent structure 
of the cloud itself or of the surrounding ionized medium. It is,
however, difficult to construct a detailed model of how the turbulence is
excited in the cloud or penetrates into it since the theory of compressible 
turbulence is still in its infancy. However, the spectrum of a homogeneous, 
compressible turbulence has been investigated in some detail and 
found to be close to the Kolmogorov one \citep{b02,bnp1,bnp2}. 
This should suffice for our qualitative estimate. 

We denote the outer scale of such a preexisting turbulence by
$R_{\rm out}$. This value should not be confused with the outer
scale of turbulence excited inside expanding HII regions, which
we analyzed in \S~\ref{difference}.  We assume that the density
fluctuations on this scale are $\delta n_{\rm out}\sim n_0\sim
10^2\,\mbox{cm}^{-3}$~\citep[e.g.,][]{larson}.  To simplify the
analysis, we neglect the interplay of turbulent fluctuations
with the boundaries of the ionized shells. We consider the
contribution of $N$~slabs of homogeneous turbulence, each of
width $\Delta R$, that lie along the line of sight. The
integrated density fluctuations are given by an expression
analogous to equation~(\ref{calN}), $\Delta {\cal N}(y)\sim
n_0(y/R_{\rm out})^{(1+\alpha) /2} {R_{\rm out}}^{1/2}(N\Delta
R)^{1/2}$.  When the interstellar ionizing radiation is not very
strong ($J\sim J_0$), the boundaries are not fully ionized and
ion-neutral collisions are important. One can assume that
turbulent fluctuations are smoothed out  
below the collisional cutoff
scale $\lambda_{\rm mfp}\sim 1.5 \times 10^{15}/n_0 
\sim 10^{13}\,\mbox{cm}$, which is assumed to be dominated by
collisions with helium atoms \citep[e.g.,][]{lithwick}.  This
estimate indicates that we can safely assume that the turbulence
is predominantly in the inertial range (with $\alpha \approx
2/3$) on the scales of interest to us.

Observations have revealed that the scaling of pulse broadening with 
dispersion measure  
changes from $\tau\propto DM^4$ for distant pulsars to $\tau
\propto DM^2$ for nearby ones. The transition between the two
scaling behaviors occurs at $DM\sim 30 \,\mbox{pc cm}^{-3}$
\citep[see, e.g., Fig. 4 in][]{bhat} and appears as 
an ``elbow'' in the $\tau$~vs.~$DM$ diagram.
This dispersion measure corresponds to 
a pulsar distance $d\sim 1\,\mbox{kpc}$, which, by the arguments
given in~\S~\ref{difference},
implies a density fluctuation scale $y\sim 10^{10}\,{\rm cm}$.
We propose that this change in the scaling behavior is a reflection of the
fact that, for nearby pulsars, the scattering angle is small and the 
scattering is dominated by turbulence within the ionized cloud boundaries 
rather than by the shape of these boundaries. Based on the arguments 
presented in \S~\ref{motivation}, the correlation scale $y$ of transmitted 
waves due to turbulent fluctuations satisfies the condition 
$\Delta \Phi(y)=\lambda  r_0\Delta {\cal N}(y)\sim 2\pi$.
Setting $\lambda=30\,\mbox{cm}$, $n_0\sim 10^2\,\mbox{cm}^{-3}$, and $y\sim  
10^{10}\,\mbox{cm}$, we infer a value of $R_{\rm out}\sim 100\, {\rm pc}$ for 
the outer scale of the turbulence. It is gratifying to note that this 
estimate agrees with the value commonly attributed (on both
theoretical and observational grounds) to the general interstellar medium (ISM) turbulence
\citep[e.g.,][]{armstrong}.   

In the case of a strong turbulence inside the boundaries ($J\gg J_0$) 
it is conceivable that the properties of the 
turbulence itself may play a key role in establishing the
non-Gaussian nature of the scattering for distant pulsars (the first scenario
outlined in \S~\ref{introduction}). In \S~\ref{origin} we
consider the possibility that molecular clouds that are situated
in the vicinity of a bright star or near the Galactic center (so
that $J\gg J_0$) may naturally develop turbulence that is strong
and non-Gaussian on small scales. We caution, however, that the
properties of the envisioned turbulence are not yet well
established. Furthermore, to fully analyze this case, one 
needs to specify how the turbulence is generated and how it interacts 
with the boundaries of the ionized layer. The answer to both of
these questions may depend on the particular mechanism of cloud
formation, which may not be universal and which, in any case, is
beyond the scope of the present discussion. Therefore, in
contrast to the robust and general conclusions that we reached
in this section on the basis of purely geometric arguments,
the results we obtain in \S~\ref{origin} are at this point still tentative.

\section{The Onset of Turbulence in Strongly Ionized Cloud Boundaries}
\label{origin}

We consider only the extreme case, when a bright star is turned on very
close to (or, alternatively, inside) a molecular cloud. In this case the
surrounding gas gets ionized up to the Str\"omgren radius, $R_s$, which
can be evaluated from the ionization balance condition $S=\frac{4}{3}\pi
R_s^3 n_2^2 \beta_2$ (where, again, $n_2$ is the density of ionized gas, 
 $\beta_2$ is the recombination coefficient, and spherical
symmetry is assumed for simplicity).  During the initial
phase of fast ionization the gas density does not change ($n_2\sim n_0$)
and one obtains $R_s\sim 2\times 10^{20} n_0^{-2/3}\,\mbox{cm}$ for
$S=10^{49}\,\mbox{s}^{-1}$, which for $n_0\sim 10^2\,\mbox{cm}^{-3}$
yields $R_s\sim 10^{19}\,\mbox{cm}$.
The duration of the fast ionization phase is $\sim 10^3-10^4\ {\rm yr}$
from the turn-on time of the ionization source
\citep[e.g.,][]{spitzer,dyson}. The pressure imbalance between the gas
inside the Str\"omgren sphere (which is rapidly heated to $T\sim
8000\,\mbox{K}$) and the surrounding cold molecular gas ($T\sim 30\,
{\rm K}$) then leads to an expansion of the ionized region, which
proceeds at approximately the sonic speed $C_2\sim
10^6\,\mbox{cm/s}$. This ``intermediate'' (or sonic expansion) phase
lasts for $\sim 10^6-10^7\, {\rm yr}$, until pressure balance
with the ambient gas is restored at the final phase of the H II region's
evolution.
{
\begin{figure} [tbp]
\centerline{\psfig{file=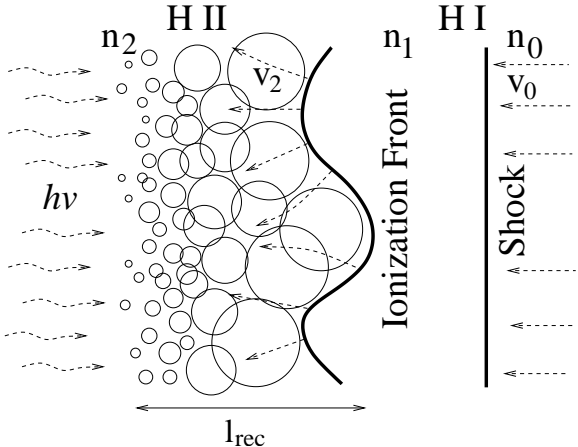,width=3.in,angle=0}}
\caption{Sketch of the structure of an unstable ionization front. The
width of the turbulent  region is of the order of the recombination
length in the ionized gas,~$l_{\rm rec}$. 
The velocity of the uncompressed
neutral gas in the reference frame of the shock is 
denoted by~${\bf V_0}$, and that of the ionized gas behind the shock
by~${\bf V_2}$. For a D-critical front 
both velocities are of the order of the sound 
speed in the ionized gas,~$C_2$.}
\label{if}
\end{figure}
}

As the lifetime of the ionizing star is typically $\sim 10^6\,
{\rm yr}$, only the intermediate phase is relevant in practice.
This phase nominally starts when the ionizing flux that reaches the
boundary of the HII region drops below 
the critical value $2 n_0 C_2$. After this time a
hydrodynamic shock forms ahead of the ionization front (IF). This shock
compresses the neutral gas to a density $n_1$ such that a weak D-type
IF is set up~\citep[e.g.,][]{spitzer}. In this
configuration (shown schematically in Fig.~\ref{if}) the speed of the
IF with respect to the compressed gas is smaller than the
D-critical speed $V_D= C^2_1/(2C_2)$, where $C_1\sim 3\times
10^4\,\mbox{cm/s}$  is the sound speed in the neutral gas. This is
equivalent to the condition $J/n_1\lesssim V_D$, which implies that the
neutral gas is compressed to $n_1\gtrsim J C_2/C^{2}_1$, which is $\gtrsim
10^5\,\mbox{cm}^{-3}$ for $J>J_0$.

In the limit when the relative speed of the IF and the
compressed gas equals $V_D$, the ionized gas flows out of the front at
the sound speed $C_2$. Since this is also roughly the speed at which the
front propagates into the (stationary) ambient gas, one can argue that
the front should always be close to the D-critical state
\citep{kahn54}. Qualitatively, when the ionizing flux reaching the front
drops below its D-critical value, the density of the ionized gas behind
the moving front drops too. Since the photon absorption is proportional
to $n^2_2$, this would lead to an increase in the flux reaching the
IF, which would turn it back into a D-critical front. Modeling
and numerical simulations confirm that the speed of the ionized gas
leaving the IF is indeed near sonic, or possibly smaller
by a factor of a few \citep{dyson,williams}.

Weak or critical D-type IFs are subject to an instability with a
linear growth rate $\sim k C_1$ (for perturbations with
wavenumber $k$), so the shortest modes initially have the
fastest growth \citep{kahn58,vandervoort,axford,williams}. The
range of scales of unstable perturbations in real fronts is
limited at the lower end by the front's width $l_{\rm if}\sim
1/(\alpha_0 n_1)$, which is $\lesssim 10^{12}\,\mbox{cm}$ in the
case we consider, and at the upper end by the recombination
length $l_{\rm rec}\sim C_2/(\beta_2n_2)$, which for our
fiducial parameters is $\sim 5\times 10^{16}\,\mbox{cm}$.
Numerical simulations by \citet{williams} have confirmed that
this instability indeed develops and evolves into the nonlinear
regime. It was found that an unstable IF becomes rough, with the
perturbation wavelength and the perturbation amplitude
saturating below the recombination length. We conjecture
(following the original speculation by \citealt{axford}) that
the IF instability would generate strong small-scale
turbulence. One can, in fact, argue quite generally that, in
analogy with a flow through a grid, a large--Reynolds-number
outflow from a rough surface should be turbulent. The numerical
simulations of \citet{williams} have indeed revealed the
existence of large velocity shears on small scales, but the
resolution of these experiments was not high enough to study the
details of the indicated fine-scale structures. We note,
however, that a compressible flow converging on a scale $l$ will
in general create density fluctuations of the order $\Delta
n_2/n_2\sim (\Delta V/C_2)^2/2$ on the same scale. For the
near-sonic conditions characterizing the downstream IF flow, the
density fluctuations could thus be close to (within a factor of
a few of) $n_2$.  A schematic picture of the envisioned
turbulence is shown in Figure~\ref{if}.

We propose that the turbulence likely excited behind IFs that
propagate into strongly ionized clouds could generate the
small-scale density perturbations inferred in the L\'evy model
of interstellar scintillations. A strong ionizing flux gives
rise to a D-type front that produces a significant compression
of the upstream neutral gas, and this in turn leads to a
sufficiently large value of the ratio $l_{\rm rec}/l_{\rm
if}\propto n_1/n_2$ for the turbulence to develop. (As we
estimated in \S~\ref{shells}, the ratio of the thickness of the
ionized skin of a molecular cloud to the mean free path of an
ionizing photon is only $\lesssim 10$ in the case of the typical
interstellar flux $J_0$.) The estimated outer scale of this
turbulence, $l_{\rm out}\sim 10^{12}-10^{16}\,\mbox{cm}$, is
much smaller than the $\sim 10\, {\rm pc}$ scale that
characterizes the turbulent H II regions invoked in the standard
interpretation of interstellar scintillations (see
\S~\ref{difference}). However, if (as can be plausibly expected) the
magnitude of the velocity fluctuations on the outer scale is
$\Delta V\lesssim C_2$ also in this case then the perturbations
should be much stronger on the scales $\sim 
10^8-10^{10}\, {\rm
cm}$ of interest. Furthermore, in the present interpretation the
turnover time of the largest eddies ($t_{\rm out} \sim
10^{6}-10^{10}\,\mbox{s}$) is typically shorter than the cooling
time $t_{\rm cool}$, and
since the ratio of the turnover time to the cooling time
decreases with the eddy scale, the difficulty associated with
the expected damping of the compressible and entropy modes  
when these two times become comparable (see \S~\ref{introduction})
would be avoided.

To illustrate the above points, we estimate the expected density
fluctuation on the small scale $y_g \sim 10^8\,\mbox{cm}$. 
The density fluctuations at the outer scale are $\Delta n_2
\lesssim 10^2\,\mbox{cm}^{-3}$. Assuming a Kolmogorov scaling
(as proposed in the \citealt{goldreich} turbulent cascade model),  
$\delta n(y)/n\sim (y/l_{\rm out})^{1/3}$.  Adopting an outer
scale of $\sim 10^{15}\,\mbox{cm}$ (for which radiative damping
should be unimportant), we infer $\delta n(y_g)\lesssim 0.1\,
\mbox{cm}^{-3}$. For comparison, the value obtained by adopting
$l_{\rm out} \sim 10\, {\rm pc}$ is a factor of $\sim 30$ smaller,
and, even more significantly, in the latter case the fluctuation
might be strongly damped by radiative cooling.

In the proposed picture turbulence is continuously regenerated
in the layer (of width $l_{\rm rec}$) behind the IF. The energy
comes from the ionizing radiation and is deposited into
fluctuating streams of hot gas that leave the rough surface of
the front. Without such regeneration, turbulence would decay on
the relatively short crossing time~$l_{\rm rec}/C_2$. 
It is also worth noting that the outer scale of this turbulence
in not the effective Str\"omgren radius (i.e., the thickness of
the ionized cloud boundary), as is sometimes assumed
\citep[e.g.,][]{yusef-zadeh} but rather the recombination 
length in the ionized gas, which is much shorter. 

To account for the inferred non-Gaussian nature of the density
fluctuations we make a connection with the fact that in the
L\'evy-flight interpretation the fluctuations responsible for
the observed radio-pulse broadening span a narrow range of 
scales, $10^8-10^{10}\,\mbox{cm}$ (see \S~\ref{difference}),
which is close to the estimated scale of the proton gyroradius $\rho_i$
in the ionized regions of molecular clouds (see
\S~\ref{introduction}). Now, the turbulent MHD cascade discussed
above corresponds to
shear Alfv\'en waves only on scales $\gg \rho_i$. On scales that
approach the ion gyroradius the Alfv\'en waves become kinetic
and develop an electric field component that is parallel to the
magnetic field. The electrons, as the most mobile particles, 
rapidly respond to this field by moving along the magnetic field
lines in the direction of increasing electric potential. 
This provides a mechanism for coupling density fluctuations in
the plasma to the cascade of shear Alfv\'en waves.
One can show that the coupling is strongest for transverse
wavenumbers $k_\perp$ satisfying $k_{\perp}\rho_i \sim 1$ and
that the Fourier amplitude of the induced 
density fluctuations may be estimated from
$\delta n_k/n\sim e\phi_k/k_B T\sim (k_{\perp}\rho_i) \delta B_k/B_0$, 
where $\phi$ is the fluctuating electric potential, $T$ is the
temperature, and $k_B$ is Boltzmann's constant~\citep[e.g.,][]{biskamp}.

The above coupling mechanism provides yet another means of
accounting for the inferred presence of strong density
fluctuations on small scales. Even more significant, however, is
the fact that, as the ion gyroradius scale is approached, the
turbulence itself gets modified. This happens because in the
presence of relatively strong electron density fluctuations 
and of the corresponding perpendicular electric field (see the above 
estimate of $\delta n_k/n$), the ion polarization drift
velocity becomes comparable to the velocity of fluid
fluctuations, and so ions can drift compressively across the
magnetic field lines. This leads to the coupling of density 
fluctuations to the Alfv\'enic fluctuations; a similar effect 
is at work in the so-called
electromagnetic drift wave (or drift Alfv\'en) turbulence, which
has been investigated in the context of plasma fusion devices
\citep[e.g.,][]{hasegawa-mima,hasegawa-wakatani,terry-horton,hazeltine,
biskamp,krommes}. Interestingly, numerical simulations of such 
turbulence have revealed a rather high level of density
fluctuations (in equipartition with kinetic fluctuations) and
statistical properties that are clearly non-Gaussian
\citep[e.g.,][]{terry,graddock}.\footnote{We are grateful to 
Paul Terry for drawing our attention to these numerical results.}

On the basis of the foregoing results one could plausibly 
attribute the generation of strong, non-Gaussian density
fluctuations in turbulent, ionized molecular gas to the onset of
compressive effects in a shear Alfv\'en wave cascade that approaches
the proton gyroradius scale. (As we discussed at the beginning
of this section, IF instability at the boundaries of molecular
clouds provides a likely mechanism for generating such
turbulence that might by itself produce strong fluctuations on
small scales.) More definitive statements must, however, await
further studies of this turbulence in an explicitly
astrophysical context.
In the meantime one could perhaps look for a signature of the
compressive effects in the density spectrum inferred from the
pulsar scintillations data on 
scales $\sim
10^8-10^{10}\,\mbox{cm}$. A likely signature would
be a flattening of the spectrum on these scales: a flattening of this
type was observed in the electron density spectrum of the solar wind
\citep{coles-harmon}, where it was, in fact, attributed to kinetic
Alfv\'en wave effects \citep{hollweg}. 

\section{Conclusion}
\label{conclusion}
To conclude our discussion, we recapitulate the main ideas 
and results of this work. We have been motivated by a recent comparison  
of radio signals from distant pulsars with an analytical 
model \citep{boldyrev1}, from which a non-Gaussian power-law
distribution of wave-scattering angles was inferred (corresponding to
L\'evy statistics). It was also deduced that the density fluctuations
responsible for this scattering should  be rather strong on small scales
($\sim 
10^8-10^{10}\, {\rm cm}$; see \S~\ref{difference}). These conclusions
differ from the predictions of the classical theory of scintillations,
which are based on Gaussian statistics.

The inferred non-Gaussian behavior of the scattering process indicates
that the scattering medium has a different structure than the one
envisioned in the standard model, where the angular deviations are
attributed to the cumulative effect of a large number of weak-scattering 
events along the line of sight. In contrast, we propose in
this paper that the observed scintillations are dominated by rare
scattering events in spatially intermittent density
structures. We identify these structures with thin and curved layers of
ionized gas, which could plausibly be the ionized boundaries of 
molecular clouds. 
In this picture, a ray propagating from a distant pulsar can 
experience significant scattering only when its trajectory intersects 
such a boundary.  
The angular deviation induced during such an encounter
can be rather large, so a single scattering event may dominate 
the total angular deflection. 
The trajectory of the ray therefore has the character of a L\'evy flight.

For a simple calculation, we modeled the ionized cloud boundaries as
thin spherical shells. We found that the scattering-angle 
probability density  
function predicted by this model has the asymptotic form 
$P(\Delta \theta)\propto (\Delta \theta)^{-5/3}$ 
for large angular deviations (see \S~\ref{shells}). Remarkably, this is
precisely the form predicted by \citet{boldyrev1} on the basis of a
comparison of the L\'evy model with observations. Our analysis therefore
strongly suggests that the scattering structures in the interstellar
medium are likely to possess a shell-like morphology.

The spatially intermittent 
nature of the electron-density distribution is an important 
ingredient of our theory. So far, analytical and 
numerical investigations of interstellar turbulence were 
mostly restricted to homogeneous settings with periodic 
(or otherwise simplified) boundary conditions. However, the actual
interstellar electron density distribution is not homogeneous, and geometric 
boundaries of scattering regions can play a dominant role in the 
line-of-sight integrated density fluctuations. Such boundary 
effects have been observed in aero-optical experiments, in which
an optical beam propagated through a confined turbulent
region \citep[e.g.,][]{dimotakis}.

The distant-pulsar relationship $\tau \propto DM^4$ between the
pulse time delay and the pulsar dispersion measure, for which
the above model provides a natural explanation, is observed to
change in nearby pulsars to $\tau \propto DM^2$, the expected
scaling for Gaussian turbulence. We have interpreted this change
as a consequence of the fact that, for nearby pulsars, the
scattering is dominated by turbulence within the ionized cloud
boundaries rather than by the shape of these boundaries. We have
found that the measured value of the transition $DM$ ($\sim
30\,\mbox{pc cm}^{-3}$) is consistent with the inferred value
($\gtrsim 10^{20}\, {\rm cm}$) of the outer scale of the general
ISM turbulence.

Although, as we just summarized, the non-Gaussian, L\'evy-type
statistics of scattering angles may have a purely geometric origin, we
also discussed (see \S~\ref{origin}) the situation in which the scattering
regions are strongly turbulent, so that the contribution from
turbulent density fluctuations dominates the scattering. This may happen
when a molecular cloud boundary is ionized by radiation from nearby
stars or by strong interstellar radiation in the vicinity of the Galactic
center. In this case a near-critical D-type ionization front develops
and propagates into the cloud.

Ionization fronts of this type are known to be linearly unstable. In its
nonlinear phase, this instability could lead to a strong turbulence on
scales below the recombination length $l_{\rm rec}\sim 10^{16}\, {\rm
cm}$ and above the ionization front thickness $l_{\rm if}\sim 10^{12}\,
{\rm cm}$ (where the numerical values correspond to a cloud of density
$\sim 10^2\, {\rm cm^{-3}}$). The outer scale of this turbulence is $\sim
l_{\rm rec}$, so the density fluctuations could be strong enough on
small scales to produce the inferred scattering. We contrast this
situation with the standard theory, where the outer scale of turbulence
is of the order of the Str\"omgren radius of an H II region ($R_s\sim
10^{19}\, {\rm cm}$). In this case the density is passively advected by the
Alfv\'enic cascade from the large outer scale to the small dissipative
one. This cascade has to pass through the radiative cooling scale ($\sim
10^{14}\, {\rm cm}$), where the density fluctuations could be
significantly attenuated, so by the time it reaches the much smaller
scales implicated in the scattering process the turbulence might be too
weak to account for the observed scintillations. In the case of a
turbulent ionization front the outer scale of turbulence is smaller than
the cooling scale and this problem is avoided.

We also noted that the small scales implied by the non-Gaussian
interpretation happen to be close to the ion gyroscale in an ionized
molecular cloud. The turbulent cascade would become compressible on
these scales because of kinetic effects, with the density fluctuations
becoming coupled to the Alfv\'enic ones. This is expected to produce a
flattening of the density spectrum in this range, providing an
alternative means of accounting for the relative strength of the
fluctuations on small scales. Even more tantalizingly for our contemplated
application, there have been indications from numerical simulations for
a clearly non-Gaussian behavior of the fluctuations in this case. However,
as neither the nonlinear development of the ionization front instability
nor the behavior of kinetic Alfv\'en turbulence in astrophysical contexts
have yet been studied in detail, our discussion of the properties of
turbulent scattering layers must be regarded as tentative at this stage.

\acknowledgements
We are grateful to Steven Cowley, William Dorland, Jeremy
Goodman, Carl Gwinn, John Krommes, 
Christopher McKee, Barney Rickett, Paul Terry, Farhad Yusef-Zadeh, and 
the referee Anthony Minter for useful comments and discussions. 
S.B. acknowledges 
the hospitality of 
the Aspen Center for Physics, where part of this work was done. A.K.
similarly acknowledges the hospitality of the Kavli Institute for Theoretical
Physics at Santa Barbara (where partial support under NSF grant PHY99-07949 
was provided). The work
of S.B. was supported by the NSF Center for Magnetic Self-Organization
in Laboratory and Astrophysical Plasmas at the University of
Chicago. A.K. was supported in part by a NASA Astrophysics Theory
Program grant NNG04G178G.

\end{document}